\documentstyle[twocolumn,aps]{revtex}
\begin{document}
\draft
\preprint{
\begin{tabular}{r}
DFTT 33/95
\\
JHU--TIPAC 95018
\\
hep-ph/9505394
\\
revised version
\end{tabular}
}
\title{
Neutrino Physics: Fundamentals of Neutrino Oscillations
}
\author{
C.W. Kim 
\thanks{E-mail: kim@eta.pha.jhu.edu ~ Permanent Address:
The Johns Hopkins University
}}
\address{
Department of Physics and Astronomy,
The Johns Hopkins University
\\
Baltimore, Maryland 21218, USA.
\\
and \\
Department of Physics, Korea Advanced Institute of Science and
Technology \\
Taejon 305-701, Korea
}
\maketitle
\begin{abstract}
In this lecture we review some of the basic properties of
neutrinos, in particular their  mass and 
the  oscillation behavior.
First we discuss how to describe the neutrino mass. Then, under 
the assumption that neutrinos are massive and mixed,
the  fundamentals
of the neutrino oscillations are discussed with emphasis on
subtle aspects which have been overlooked in the past. 
We then review the terrestrial neutrino oscillation
experiments in the framework of three generations of
neutrinos with the standard mass hierarchy. Finally,
a brief summary of the current status of the solar 
and atmospheric neutrino problems will be given.
\end{abstract}
 
\section{Introduction}
Neutrinos are still the least known particles among the 
well-established fundamental fermions in the Universe.
The most important property of the neutrinos is whether
or not  neutrinos are massive. 
If they turn out to be massless, one must understand
the reason why they are massless. The masslessness of the
photon is guaranteed by the local $U(1)_{em}$ gauge invariance.
Because of the lack of such a symmetry and in light of the
natural occurrence of the massiveness of neutrinos when we 
extend the standard model of strong and electroweak
interactions to accommodate several problems in the
model, it is very tempting to assume that neutrinos 
are indeed massive, although no theory can predict with any certainty
the values of neutrino masses.
 In this lecture,   we begin with a brief
summary of  the  current status
of many efforts to find the neutrino mass and how to 
describe the mass \cite{kim} \cite{bohm} \cite{pal}.   

The direct mass measurements based on the decay
kinematics of $ ^3 {\mbox{H}}$, 
$\pi$ and $\tau$ have steadily been improving but
 it now appears that we have reached the end of
the road on this efforts unless  a new generation
of the techniques is forthcoming. 
The current limits are
\begin{eqnarray}
m(\nu_{e})& \lesssim & 4.5~~ \mbox{eV} \\ \nonumber
m(\nu_{\mu}) &  \lesssim & 160~ \mbox{KeV} \\ \nonumber
m(\nu_{\tau}) & \lesssim & 24 ~~\mbox{MeV}.
\end{eqnarray}
It is interesting to remark that in the cases of 
both
$m(\nu_{e})$
and
$m(\nu_{\mu})$,
all the experimental analyses
have consistently yielded $negaive$ $m^2$,
which shows that some systematics in experiments are
still not properly understood.

On the other hand we have very interesting limits
from cosmology. Requiring that neutrinos cannot 
over-close the Universe, we have 
\begin{equation}
\sum_{i} m_{i}  \lesssim (20 \sim 30) \mbox{eV},
\end{equation}
where the uncertainty is due to that of
the Hubble constant. The above limit applies
only when neutrinos are stable or much longer lived than
the age of the Universe.
The recent COBE data and other 
astronomical observations provide us with an intriguing
possibility that the heaviest neutrino mass might be
of oder of 1 $\sim$ 10 eV. A neutrino or neutrinos of this mass
 range can  explain the small scale region of
the power spectrum in the form of
hot dark matter.
Currently, however, the most promising way to probe
the neutrino mass is considered to be the  neutrino
oscillation experiments.  The source can be either
terrestrial or extraterrestrial.

Indications of massive neutrinos  from the
oscillation experiments with reactors and accelerators
have not been  found for  many decades with an exception
of the recent LSND experiment \cite{LSND}.
We will critically analyze all the reactor and accelerator
oscillation experiments using three generations
of neutrinos with the standard mass
hierarchy. We show  that
in contrast to the simple two generation analysis,
the LSND allowed
region lies in the forbidden region of all the previous
terrestrial oscillation experiments,  when the previous data 
are analyzed     
 in the framework of three generations of
neutrinos with the standard  mass hierarchy. 
Also discussed are implications of the three 
generation analysis of neutrinoless $\beta$ $\beta$ decay.   
Finally, we will briefly review the present status
of the solar and atmospheric neutrino problems.

\section{Neutrino Mass}
In the past, there have appeared numerous papers
on the theory of neutrino mass. For recent references,
see, for example, \cite{kim}, \cite{bohm} and \cite{pal}.
None of theories,
however, is 
satisfactory though each theory has its own merit.
Basically, there are two ways to generate
neutrinos masses. First, one modifies the Higgs
sector in the standard model.
For example, an additional singlet, doublet or triplet
with or without right-handed neutrinos is added to
the original Higgs doublet in the standard model.
In this case, however, one is forced to introduce
a new mass scale in the form of the vacuum expectation
value. This, however, is not an explanation of the small
neutrino mass. The other possibility is to utilize
extremely heavy
right-handed neutrinos which 
appear in models such as left-right symmetry
models or GUTS.
In the following we list several ways to $describe$  
the neutrinos mass \cite{kim} \cite{samoil},
eventually leading to  detailed discussions of
the second possibility.
 
\begin{enumerate}
\item {\bf{Dirac Mass}.}
The simplest way to describe the mass is, of course,
 to introduce the right-- handed neutrinos although
they have not been seen so far, at least in the 
energy region accessible to us now.
The mass term in the Lagrangian is
\begin{equation}
{\cal{L}}_{Dirac} = - 
({\overline{ \nu}}_{L} {\cal{M}} 
\nu_{R} + {\overline{\nu}}_{R}
{\cal{M}}
^{\dagger}
\nu_{L}),
\end{equation}
where
$\nu_{L,R}$ is given by
\begin{equation}
\nu_{L,R}=\left(
	\begin{array}{l}
	\nu_e \\
	\nu_\mu \\
	\nu_\tau 
	\end{array} \right)_{L,R}.
\end{equation}
In general, 
${\cal{M}}$ 
is a $3 \times 3$ complex
mass matrix, and hence there is no
guarantee that mass eigenvalues are
positive.
One needs to bi-diagonalize 
${\cal{M}}$
using two unitary matrices
$U$ and $V$:
\begin{equation}
U^{\dagger}
{\cal{M}}
V = m_{D} =
\left( \begin{array}{lll}
m_{1} & 0 & 0 \\
0 &   m_{2} & 0\\
0 & 0 & m_{3}
\end{array}\right)~,
\end{equation}
where $U$ and $V$ relate
the mass eigenstates
$\nu^{(m)}_L$
 and weak eigenstates
$\nu_L$ as
\begin{eqnarray}
\nu_{L} & = & U \nu_{L}^{(m)}\\ \nonumber
\nu_{R} & = & V \nu_{R}^{(m)}.
\end{eqnarray}
The diagonalized mass Lagrangian is
\begin{equation}
{\cal{L}}_{Dirac}= -
{\overline{\nu}}_{L}^{(m)} m_{D} \nu_{R}^{(m)}
+ \mbox{h.c.}.
\end{equation}
Physically, of course, since only $\nu_L$ is involved
in weak interactions, the $U$ is the Cabibbo, Kobayashi
and Maskawa mixing matrix. In this case, the basic 
questions to be answered would be 
where $\nu_R$ is and why $m_i$   are  so small.
\item {\bf{Majorana Mass}.}
The Majorana neutrino mass can be described by
the use of $\nu_L$ alone:
\begin{equation}
{\cal{L}}_{Majorana} = -\frac{1}{2} 
{\overline{ \nu^{C}}}_{L} {\cal{M}} 
\nu_{L} + \mbox{h.c.},
\end{equation}
where we note that $\nu^{C}_{L}$ is
a right-handed neutrino.
Since
\begin{equation}
\overline{ \nu}^C_{L} {\cal{M}} 
\nu_{L} =
\overline{ \nu}^C_{L} {\cal{M}}^{\mbox{T}} 
\nu_{L},
\end{equation}
we have ${\cal{M}} = {\cal{M}}^{\mbox{T}}$, i.e.
${\cal{M}}$ is symmetric and diagonalization can be
done by  a single unitary matrix $U$ in this case.
Again, with 
\begin{equation}
\nu_{L} = U \nu^{(m)}_{L},
\end{equation}
Eq.(8) becomes
\begin{equation}
{\cal{L}}_{Majorana} = - {\frac{1}{2}} 
[{\overline{ \nu^{C}}}_{L} {m_D} 
\nu_{L}  +
{\overline{ \nu}}_{L} {m_D} 
\nu^{C}_{L}].
\end{equation}
Defining
 \begin{equation}  
\nu_{Maj} \equiv \nu_{L} + \nu^{C}_{L},
\end{equation}
which is  clearly Majorana neutrino, we can rewrite
\begin{equation}  
{\cal{L}}_{Majorana} = - {\frac{1}{2}} 
{\overline{ \nu}}_{Maj} {m_D} 
\nu_{Maj}.
\end{equation}
Although the expression of Eq.(13) looks
similar to that of Eq.(7),  there is a fundamental
difference between them. In Eq.(13), the lepton
number is  violated by two units.
\item {\bf{Dirac--Majorana Mass}.}
First, let us consider the one generation case.
The Lagrangian of interest is
\begin{equation}
{\cal{L}}_{D-M}=-M \bar{\nu}_L \nu_R -
	\frac{1}{2} \left(m_L \bar{\nu}^C_L \nu_L +
			m_R \bar{\nu}^C_L\nu_R \right)
	+ h.c.~,
\end{equation}
where $M$ is Dirac mass and $m_{L}(m_R)$ are
Majorana masses.
If we define a left--handed neutrino state $\nu$ as
\begin{equation}
\nu \equiv {\nu_L \choose \nu_R^C }~,
\end{equation}
the Dirac--Majorana Lagrangian looks like
that of  Majorana  :
\begin{equation}
{\cal{L}}_{D-M} = -{1 \over 2} \bar{\nu}^C {\cal{M}}\nu
		+ h.c.~,
\end{equation}
where the mass matrix ${\cal{M}}$ is
\begin{equation}
{\cal{M}}={
m_L ~ M \choose
M ~ m_R}~.
\end{equation}
It is to be noted that the state $\nu$ is not a 
mass eigenstate.
Diagonalizing ${\cal{M}}$ yields
\begin{eqnarray}
m_1 &=& {1 \over 2}\sqrt{4M^2+(m_R-m_L)^2}
		-\frac{m_L+m_R}{2}
	\\	\nonumber
m_2 &=& {1 \over 2}\sqrt{4M^2+(m_R-m_L)^2}
		+\frac{m_L+m_R}{2}~.
\end{eqnarray}
Now, the mass eigenstate $\nu^{(m)}$ can be defined as 
\begin{eqnarray}
\nu^{(m)} &=& {\nu_1 \choose \nu_2} =U\nu
	\\ \nonumber
    &=& {  {\cos\theta \nu_L-\sin\theta \nu_R^C}
		\choose
           {\sin\theta \nu_L+\cos\theta \nu_R^C} }~,
\end{eqnarray}
where the mixing angle is 
given by  $\tan 2 \theta =2M/(m_R-m_L)$.

Now let us consider  two interesting cases.
\begin{enumerate}
\item{Case with $M \gg m_L, m_R$.}
In this case $m_1$ and $m_2$ are almost degenerate in mass
( Eq.(18) implies $m_1 \simeq m_2 \simeq M$) and
we have $\theta \simeq 45^o$.
This case is called special (because
 $\theta \simeq 45^o$)
 pseudo--Dirac neutrino \cite{pse}.
In this case, $\nu_1$ and $\nu_2$ have opposite CP phase.
And we have half active $\nu_L$ and half sterile $\nu_R^C$.

\item{Case  with  $m_R \gg M,m_L$.}
For simplicity, we assume  $m_L=0$.
Then, we have
\begin{eqnarray}
m_1 &\simeq& \frac{M^2}{m_R} 
\\  \nonumber
m_2 &\simeq m_R~,
\end{eqnarray}
implying that $m_1$ is naturally small and $m_2$ is large.
Since $\theta \simeq 0$, $\nu_L$ and $\nu_R^C$
 are practically decoupled.
This is the seesaw mechanism \cite{seesaw}.
Note that $M$ has 
the standard model mass scale whereas $m_R$ 
is the mass scale of the right -- handed neutrino
associated with models such as L--R symmetry, SO(10),
$E_6, \ldots$  models.

\end{enumerate}
\item {\bf{Seesaw Mechanism}.}
In order to apply the seesaw mechanism to 
practical cases, let us generalize the above to
the three generation case by writing
\begin{equation}
m_1=\frac{m_D^2}{M_R} \rightarrow
	\overline{m}_1 = \overline{m}_D \frac{1}
{ \overline{M}}
	\overline{m}_D^T~,
\end{equation}
where all barred  objects  are $3\times 3 $ matrices
and as before we assume $|(\overline{m}_D)_{ij}|
 \ll |\overline{M}_{ij}|$.
Now
the matrix analogous to Eq.(17)
is a $6 \times 6$ mass matrix $\overline{\cal{M}}$  given by 
\begin{equation}
\overline{\cal{M}}={
			0 ~~ \overline{m}_D \choose
			\overline{m}_D ~ \overline{M}_R }~.
\end{equation}

At this point we can consider two possible cases:
1). There is no mass hierarchy among the right--handed
neutrinos, i.e.

\begin{equation}
\overline{M}_{R} \simeq M_R \left(
		\begin{array}{rrr}
		1&0&0\\
		0&1&0\\
		0&0&1
		\end{array}\right)~,
\end{equation}
implying
\begin{equation}
m(\nu_1) : m(\nu_2) :m(\nu_3) =
\frac{m_u^2}{M_R}:
\frac{m_c^2}{M_R}:
\frac{m_t^2}{M_R}~.
\end{equation}
This is called the  quadratic seesaw mechanism.

2). The next possibility is the case where
the right--handed neutrinos have the mass hierarchy
similar to that of the known "top" quarks, i.e.
\begin{equation}
\overline{M} \simeq M_R \left(
		\begin{array}{rrr}
		\frac{m_u}{m}&0&0\\
		0&\frac{m_c}{m}&0\\
		0&0&\frac{m_t}{m}
		\end{array}\right)~.
\end{equation}
 In this case we have
\begin{equation}
m(\nu_1) : m(\nu_2) :m(\nu_3) =
\frac{m_u}{M_R}:
\frac{m_c}{M_R}:
\frac{m_t}{M_R}~.
\end{equation}
which is called  the linear seesaw mechanism.

The above relations as given by Eqs.(24) and (25)
are valid at the GUTS scales which means one has to
bring them down to the low energy region of our interest
using the Renormalization Group Equations (RGE). 
Considering only the running of the mass in one--loop
calculations, for example , Eq.(24) is modified as
\cite{RGE}
\begin{eqnarray}
& & 
m(\nu_1) :m(\nu_2): m(\nu_3)
\\ \nonumber
&=&\left\{
\begin{array}{ll}
 0.05 \frac{m_u^{2}}{M_R} :
 0.09 \frac{m_c^{2}}{M_R} :
 0.38 \frac{m_t^{2}}{M_R} & \mbox{SUSY SU(5)}
\vspace{0.3 cm}
\\
 0.05 \frac{m_u^{2}}{M_R} :
 0.07 \frac{m_c^{2}}{M_R} :
 0.18 \frac{m_t^{2}}{M_R} & \mbox{SO(10)}.
\end{array}\right.
\end{eqnarray}
As we can see above, the corrections due to
the RGE depend on the models of choice.
In the seesaw mechanism, the actual size of 
neutrino masses will be determined by
 the mass scale for $M_R$.
In GUTS such as SUSY SU(5), we have
\begin{eqnarray}
M_R(\mbox{mass of $\nu_R$}) &\sim& h V_{GUT} 
	\sim \left({h \over g}\right) M_X,
	\\ \nonumber
M_X &\sim& gV_{GUT},
\end{eqnarray}
whereas for the SO(10) model, we have
\begin{equation}
M_X \sim
 \left({h \over g}\right) M_{L-R}.
\end{equation}
In the above, $h$, $V_{GUT}$,  and $g$ are,
respectively, the Yukawa coupling, GUTS vacuum
expectation value and the gauge coupling constant.
It is generally expected that the mass scale of $M_R$ is
$10^{10}$ GeV $\lesssim M_R \lesssim 10^{15}$ GeV
, naturally explaining the smallness of the neutrino mass.
\end{enumerate}

\section{Neutrino Oscillations}
If neutrinos are massive and mixed,
neutrinos are produced and detected  in the form 
of the weak eigenstates whereas  when they propagate
from the point of the production to their detection,
their motion is dictated by the mass eigenstates.
This   leads to the phenomenon of  neutrino
oscillations \cite{maki}. For neutrino oscillations to
ocurr, neutrinos must be massive and mixed. 
The weak eigenstates and mass eigenstates are
related   by a unitary matrix $U$ as
\begin{equation}
\nu_W=U\nu_M ~,
\end{equation}
where $U$ is parameterized by,  in the case of 
the two generations of neutrinos,
\begin{equation}
U={~\cos\theta ~~\sin\theta \choose -\sin\theta~~\cos\theta}~.
\end{equation}
In the above it was assumed that neutrinos are
Dirac. For Majorana neutrinos,  a CP phase   may appear.
The equation of motion for the mass eigenstate is 
\begin{equation}
i{\dot{\nu}_M}=H\nu_M =
{E_1 ~~0 \choose 0 ~~E_2}\nu_M ~~;~\nu_M={\nu_1 \choose \nu_2}~.
\end{equation}
Since neutrinos are expected to be extremely 
relativistic,
using the following approximation
\begin{equation}
E_i=\sqrt{p^2+m_i^2}\simeq p+\frac{m_i^2}{2p}
~:~ p\simeq E~,
\end{equation}
we have from Eq.(32)
\begin{equation}
i\dot{\nu}_M = 
\left[ E \cdot 1 + \frac{1}{2E}
		{m_1^2 ~0 \choose 0 ~ m_2^2} \right] \nu_M~.
\end{equation}
Replacing $\nu_M \rightarrow e^{i\alpha t}\nu_M$,
the above equation becomes
\begin{equation}
i\dot{\nu}_M=(E+\alpha)\nu_M +\frac{1}{2E}
	{\cal{M}}_M
\nu_M~,
\end{equation}
where ${\cal{M}}_M=diag(m_1^2,m_2^2)$.
Taking $\alpha=-E$, we have
the equation of motion for $\nu_M$ as 
\begin{equation}
i\dot{\nu}_M=
\frac{1}{2E}
	{\cal{M}}_M
\nu_M~.
\end{equation}

Now, using $\nu_M=U^\dagger(\theta)\nu_{W}$(Eq.(30)),
the equation of motion for weak eigenstates can be derived as
\begin{equation}
i\dot{\nu}_M = 
\frac{1}{2E} 
U(\theta){\cal{M}}_M U(\theta)^\dagger 
\nu_W~,
\end{equation}
where 
\begin{eqnarray}
{\cal{M}}_W &\equiv& 
U(\theta){\cal{M}}_M U(\theta)^\dagger 
\\ \nonumber
&=& \left( \begin{array}{ll}
m_1^2 \cos^2\theta + m_2^2\sin^2\theta 
& \frac{\Delta}{2}\sin 2 \theta
\\
\frac{\Delta}{2}\sin 2 \theta &
m_2^2 \cos^2\theta + m_1^2\sin^2\theta 
\end{array} \right)~. 
\end{eqnarray}
In Eq.(38),
\begin{eqnarray}
\Delta &\equiv& m_2^2-m_1^2
\\ \nonumber
\langle m(\nu_e)^2\rangle &\equiv& 
m_1^2 \cos^2\theta + m_2^2\sin^2\theta 
\\ \nonumber
\langle m(\nu_\mu)^2\rangle &\equiv& 
m_2^2 \cos^2\theta + m_1^2\sin^2\theta. 
\end{eqnarray}
By solving the above differential equation, 
one can easily obtain the well--known oscillation
probability for the two generation case. For example, 
\begin{equation}
P(\nu_\mu \rightarrow \nu_e) =
 \sin^{2}(2\theta)\sin^{2}(\Delta L /4 E_{\nu}).
\end{equation}
A similar treatment for the three generation case 
can be trivially performed but it is not very 
informative. Since we shall discuss three generation
cases below,  we close this section with the  
 case of two generations.
  
\section{Wave Packet Effect}
 The above treatment of the neutrino oscillations
assumes explicitly that the neutrinos that are produced
via weak interactions are described by plane waves.
However, in practice
since neutrino--emitting particles are never in a  free stable
state,
neutrinos cannot be described by  plane waves.
Rather, they  must be described by wave packets,
 due to pressure broadening or collisions of neutrino--emitting
particles \cite{wave}.
Therefore, the size of wave packet $\sigma_x$ is 
given by dimension of the
region within which production processes are effectively localized
\begin{eqnarray}
\sigma_x &\sim& \frac{\mbox{mean free path }(l)}
	     {\mbox{mean thermal velocity}(v)}    
\\ \nonumber
&\sim& \frac{T^{3\over 2}}{N},
\end{eqnarray}
 where we have used
\begin{eqnarray}
l & \sim & \frac{T^2}{N} \\ \nonumber
	v &\sim & T^{1\over2}.
\end{eqnarray}
It is easy to see that $\sigma_x \rightarrow \infty$
as $l \rightarrow \infty$ and $v \rightarrow0$,
recovering the case of plane waves. 
Let us consider several examples in order to see how
large or small the size of $\sigma_x$ is.
\begin{enumerate}
\item{\bf {Solar Core}.}
For neutrinos emitted from the process
$^8 B \rightarrow ^8 Be^* + e^+ + \nu_e$, we
have typically $T\sim 1.3$ KeV and $\rho \sim 120 g/cm^3$,
implying $\sigma_x \sim 10^{-7} cm$.
For neutrinos emitted from the process
$p+p \rightarrow ^2 H + e^+ + \nu_e$,  one has
typically $T\sim 1.1$ KeV and $\rho \sim 100 g/cm^3$,
leading to 
$\sigma_x \sim 5\times 10^{-7} cm$.

\item{\bf{Nuclear Reactor}.}
In the case of neutrinos from nuclear reactors, we  have 
$v \sim P_N / M_N \sim 10^{-5}$,
$N \sim 10^{23} /cm^3$ and $l_N \sim 10^{-7} cm$,
leading to 
$\sigma_x \sim 10^{-4} cm$.

\item{\bf{Accelerator/Atmospheric Neutrinos}.}
Since they are decay--products of $\pi$, or $\mu$,
$\sigma_x$ is characterized by the typical
weak interaction length, $t_W$, so that
$\sigma_x \sim ct_W \gtrsim 10^{2} cm$.

\item{\bf{Supernova Neutrinos}.}
Neutrinos from the core with $T\sim 10$ MeV 
and $\rho \sim 10^{14} g/cm^3$
have $\sigma_x \sim  10^{-13}cm$.
Neutrinos from the neutrino sphere with $T\sim 1$ MeV and 
$\rho \sim 100 g/cm^3$ have $\sigma_x \sim 10^{-9} cm$.
\end{enumerate}
 
What do wave packets do to neutrino oscillations?
Recalling that a plane wave is described by
$|\psi_a(x,t)>=e^{ipx-E_at}|\nu_a>$
with $E_a=\sqrt{p^2+m_a^2} \simeq p+m_a^2/2p$,
 and using the wave packet in the momentum space
\begin{equation}
\psi_a(p)
=({\frac{1}{\sqrt{2\pi}\sigma_p}})^{1\over 2} 
e^{-\frac{(p-p_a)^2}{4\sigma_p^2}},
\end{equation}
the wave packet in the space is described by, with 
$ \sigma_p \sigma_x \sim 1$,
\begin{equation}
|\psi_a(x,t)>
=\left(\frac{1}{\sqrt{2\pi}\sigma_p}\right)^{1\over 2} 
e^{   ipx-E_at-
\frac{(x-v_a t)}{4\sigma_x^2}
	}|\nu_a>.
\end{equation}
By using this wave packet in 
a way similar to the standard treatment of
neutrino oscillations,  instead of the
usual expression  for the plane wave, 
\begin{eqnarray}
P(\nu_e \rightarrow \nu_\mu)
	& \sim & \sum_{a,b}
U_{\nu_{{\mu} a}} U_{{\nu{e} b}}^{*}
U_{\nu_{{\mu} b}}^{*} U_{{e} b} 
~~e^{i\frac{2\pi L}{L_{a,b}^{osc}}}
,
\end{eqnarray}
one finds,  for wave packets \cite{wave},
\begin{eqnarray}
P(\nu_e \rightarrow \nu_\mu)
	& \sim &\sum_{a,b} 
U_{\nu_{{\mu} a}} U_{{\nu{e} b}}^{*}
U_{\nu_{{\mu} b}}^{*} U_{{e} b} 
~~e^{i\frac{2\pi L}
{L_{a,b}^{osc}}}
	e^{-\frac{L}
{L_{a,b}^{coh}}}
,
\end{eqnarray}
where
\begin{equation}
{L_{a,b}^{osc}} 
\equiv \frac{4\pi E}{m_a^2-m_b^2}
\end{equation}
is the oscillation length and
\begin{equation}
{L_{a,b}^{coh}}
\equiv 4\sqrt{2}E^2\frac{\sigma_x}{m_a^2-m_b^2}
\end{equation}
is  the coherence length  beyond which neutrinos
  do not oscillate  in contrast to the case of plane waves.
Therefore, the wave packets effects introduce
, as can be seen above,  an exponentially damping
factor which is present only when $ a \neq b $,
implying that the oscillating term would eventually
disappear beyond the coherence length. 
Typical coherence  lengths based on the above mentioned
sizes of $\sigma_x$ are:
\begin{enumerate}
\item 
{\bf{Solar Neutrinos}}:~~~~~ $L^{coh} \sim 10^{8} {\mbox{cm}}$.
\item 
{\bf{Reactor Neutrinos}}:~~~ $L^{coh} \sim 10^{9} {\mbox{cm}}$.
\item
{\bf{Accelerator Neutrinos}}:  $L^{coh} \sim 10^{21} {\mbox{cm}}$.
\item
{\bf{Supernova Neutrinos}}:  $L^{coh} \sim 100 {\mbox{cm}}$.
\end{enumerate}
The above numbers are based on typical values of
$E_{\nu}$ and $\Delta m^2 \sim 1 \mbox{eV}^2 $ 
in their respective cases. Therefore, in the usual reactor
 and accelerator
experiments,
unless $\Delta m^2$ turn out to be unexpectedly large
  the oscillation lengths are long enough
so that one can indeed observe oscillations.
Note, however,  that in the case of supernova neutrinos,
the  coherence length is so short that neutrinos from
supernova do not oscillate on the way to the Earth.
This, however, does not mean that there  occur no flavor
changes among three generations of neutrinos. They simply
do not have oscillating terms in the  transition probability.
 \section{Weak and Mass Eigenstates}
The question we like to ask in this Section is 
whether or not the weak eigenstates make sense. 
In the standard treatment of neutrino oscillations,
one always uses the weak eigenstates to discuss oscillations
in vacuum and in matter. But, here we will show that
strictly speaking, the weak eigenstates do not make sense,
although the weak eigen fields are well-defined \cite{weak}.
In order to illustrate the point, let us consider the
detection of $\nu_{e}$ via $ \nu_{e} + n \rightarrow
p + e^{-}$. The amplitude of interest is 
\begin{eqnarray}
{G_{F}\over \sqrt{2}}&\langle& e^{-}, p|
\bar{e}\gamma^\mu (1-\gamma_5)
\nu_e J_\mu^{(h)}(m_{i}) | \nu_\alpha, n \rangle
\\ \nonumber
&=& \sum_i 
U_{ei}
U_{\alpha i}^*
\left[
	\langle e^- |
		\bar{e}\gamma^\mu (1-\gamma_5) \nu_i
		    |\nu_i \rangle
		h_\mu^{(i)}(n,p)
\right]~,
\end{eqnarray}
where the equality was obtained by using
\begin{eqnarray}
\nu_{e} & = & \sum_i U_{ei}\nu_{i} \\ \nonumber
|\nu_\alpha \rangle & = & \sum_i {U_{\alpha i}}^*
|\nu_{i}\rangle.
\end{eqnarray}
The crucial observation here is that we deliberately set
$\nu_{\alpha}$ instead of $\nu_e$. We do know from the definition
of the amplitude on the left-hand side
  that it is zero unless $\alpha = e$.
However, the right hand side of Eq.(48) is not necessarily
zero even if $\alpha =e$ is  satisfied, since unless
the quantity in the square bracket is factored out from the
sum, we do not have the orthonomality
\begin{equation}
\sum_{i} U_{ei} U^{*}_{\alpha i}= \delta_{\alpha e}.
\end{equation}
The factorization is possible only when neutrinos are relativistic
and the hadron part is not affected by the presence of the neutrino
mass.
This simple observation clearly indicates that unless
neutrinos turn out to be always relativistic, a caution is needed
when one uses the concept of the  weak eigenstates.
 In fact, it has been shown that the
 Fock space of the weak eigenstates
$|\nu_{\alpha}\rangle$ does not exist \cite{weak}.
 The Fock space can be
 shown to exist only in the limit of the relativistic neutrinos.
Hence, only if and when neutrinos are non-relativistic, 
one is allowed to calculate
the standard oscillation probability without using the concept
of the weak eigenstates. A rigorous calculation
without resort to the weak eigenstates has already 
 carried out \cite{carlo}. One can demonstrate that
 a specific production and detection process of certain
weak eigenstate neutrinos can be expressed  simply by 
using the mass eigenstates, and only in the extreme
relativistic case, the well--known transition probability
can be factored out from the expression.

\section{Terrestrial neutrino Oscillations}
In this Section, we will review the latest data on 
the reactor and accelerator oscillation experiments
in the framework in which the three masses,
$m_1$, $m_2$ and $m_3$ of
the mass eigenstates $\nu_1$, $\nu_2$ and $\nu_3$ are
ordered in such a way as suggested by the most
popular mechanism  that can  explain the smallness
of the neutrino mass, i.e. seesaw mechanism \cite{seesaw},
\begin{equation}
m_1 \ll m_2 \ll m_3 .
\end{equation}
An additional assumption is that the the solar
neutrino problem is real and the deficit is at least
10 percent or more of the standard 
solar model prediction. The details on the deficits of 
 the four current experiments are  not 
important for our argument.
In any case we need  to assume 
\begin{equation}
\Delta m_{21}^{2} \lesssim 10^{-3}~~{\mbox{eV}^2}.
\end{equation}
Of course we are also assuming the standard mixing denoted by the 
unitary transformation
\begin{equation}
\nu _\alpha = \sum_{i=1} ^{3}U_{\alpha i} \nu_i, 
\end{equation}
where $\alpha = e, \nu $, and $\tau$ and $\nu_i$ is
a mass eigenfield
 with mass $m_i$. (The arguments presented here
are based on the two papers \cite{giunti1} and
\cite{giunti2}.)
The environment of the terrestrial oscillation experiments is 
summarized by the following  ranges of the distances
($L$) between the source and the detector and the neutrino
energies($E_\nu$)
\begin{eqnarray}
L & \simeq & 10 {\mbox{m}}  \sim  1 {\mbox{Km}},\\ \nonumber
E_\nu & \simeq & 1  {\mbox{MeV}} \sim 10 {\mbox{GeV}}.
\end{eqnarray}
Because of the assumption of the mass hierarchy
as assumed in Eq.(52), we have two independent values of 
$\Delta m^2$, i.e. $\Delta m_{21}^2$ and $\Delta m_{31}^2$.
For the terrestrial neutrino oscillations, we can see 
with Eq.(40) that
\begin{equation}
\mbox{sin}^{2}(\frac{\Delta m_{21}^2 L}{4 E_\nu}) \simeq 0 ,
\end{equation}
which implies that there remains only one mass scale
in the scheme, i.e. $\Delta m_{31}^2 \equiv \Delta m^2$. 

Under these assumptions, the transition probability
for the oscillation $ \nu_{\alpha} \rightarrow
\nu_{\beta}$ is given by
\begin{eqnarray}
P(\nu_{\alpha}\rightarrow
 \nu_{\beta})=~~~~~~~~~~~~~~~~~~~~~~~~~~~~~~~~~~~~~~~~~~\\ \nonumber
\left |U_{\beta 1} U_{\alpha 1} +
 U_{\beta 2}U_{\alpha 2} 
+ U_{\beta 3}U_{\alpha 3}\exp \lbrace {\frac{-i \Delta m^2_{31} L}
{2 E}}\rbrace \right |^2.
\end{eqnarray}
Note that there appears only one oscillation term
in Eq.(57). From Eq.(57), we have ,for example, 
\begin{eqnarray}
P(\nu_{\mu}\rightarrow \nu_{e}) & = &
A(\nu_\mu \rightarrow \nu_e) {\mbox{sin}}^{2}(\frac{
\Delta m_{31}^2 L}{4 E_\nu}),\\ \nonumber
P(\nu_\mu \rightarrow \nu_\tau) & = &
A(\nu_\mu \rightarrow \nu_\tau)\mbox{sin}^{2}(
\frac{\Delta m_{31}^2 L}{4 E_\nu}), 
\end{eqnarray}
where the oscillation $factors$ A's are given by
\begin{eqnarray}
A(\nu_\mu \rightarrow \nu_e) & = &
4 {\mid}U_{\mu3}{\mid}^2 {\mid}U_{e3}{\mid}^2,\\ \nonumber
A(\nu_\mu \rightarrow \nu_\tau) & = &
4 {\vert}U_{\mu3}{\vert}^2 {\vert}U_{\tau3}{\vert}^2  = 
4 {\mid}U_{\mu3}{\mid}^2 
(1-
{\mid}U_{e3}{\mid}^2
- {\mid}U_{\mu3}{\mid}^2).
\end{eqnarray}
It is to be emphasized here that in their appearance,
the oscillation probabilities given  
in Eq.(58) take the same
form as the two generation case. In fact, the two generation
cases are reproduced when $A(\nu_{\mu} \rightarrow 
\nu_e)$ and $A(\nu_\mu \rightarrow \nu_\tau)$ are replaced
by $\mbox{sin}^2(2\theta_{e\mu})$ and 
$\mbox{sin}^2(2 \theta_{\mu \tau})$,
respectively. However, physical implications are quite
different. Namely, in the case of two generations
the two probabilities have nothing to do with each other
and are not related in any way. But, in the three generation
case, two probabilities are inter-related 
since both are determined
by the same
$|U_{\mu3}|^2$ 
and
$ |U_{e3} |^2 $.

Similarly, the survival probabilities $P(\nu_\alpha 
\rightarrow \nu_\alpha)$ are given by
\begin{equation}
P(\nu_{\alpha} \rightarrow \nu_{\alpha})=
1- B(\nu_{\alpha} \rightarrow \nu_{\alpha})
\mbox{sin}^2(\frac{\Delta m^2}{4E}),
\end{equation}
where the oscillation  factor 
$B(\nu_{\alpha} \rightarrow \nu_{\alpha})$
 is given by
 \begin{equation}
B(\nu_{\alpha} \rightarrow \nu_{\alpha})
= 4 
 |U_{\alpha 3}|^2
(1 - |U_{\alpha 3}^2 |^2).
\end{equation}
Therefore, all the oscillation probabilities are
completely determined by the three parameters
\begin{equation}
\Delta m_{31}^2 \equiv \Delta m^2 ,~~ 
{\vert}U_{e3}{\vert}^2,~~ 
\mbox{and}~~
{\vert}U_{\mu3}{\vert}^2
\end{equation}
and oscillation probabilities are all 
inter-dependent. This is simply not the case
in the two generation case. Furthermore, in our model, we have
\begin{equation}
P(\nu_{\alpha} \rightarrow \nu_{\alpha})=
P(\nu_{\bar{\alpha}} \rightarrow \nu_{\bar{\alpha}}),
\end{equation}
without the assumption of CP conservation.

We now proceed to analyze the data using the formulas
derived   above. The data  to be 
 used in our analysis are as follows:
Bugey \cite {bugey}  $\nu_e \rightarrow \nu_x$ , CDHS,\cite
{CDHS} and CCFR84 \cite{CCFR} $\nu_\mu \rightarrow \nu _x$
and  BNL E776 \cite{BNL}, KARMEN \cite{KARMEN} and
LSND \cite{LSND} $\nu_{\mu} \rightarrow 
\nu_{e}$ experiments.

First, from the disappearance (or survival) experiments, we can
find limits on the oscillation amplitudes B's as
\begin{eqnarray}
B(\nu_e \rightarrow \nu_e) & \lesssim  & B^0(\nu_e),\\ \nonumber
B(\nu_\mu \rightarrow \nu_\mu) & \lesssim & B^0(\nu_\mu),
\end{eqnarray}
where $B^0$'s are obtained from the data for
fixed values of $\Delta m^2$ and are, in general, small numbers
since no clear disappearance of 
the initial neutrinos has been observed
as yet. Using Eq.(16), and solving  it for
${\vert}U_{\alpha3}{\vert}^2$, we find the following 
two bounds 
\begin{eqnarray}
U_{\alpha 3} & \lesssim &  \frac{(1-\sqrt{1+B^{0}(\nu_\alpha)})}{2}
:\mbox{small},\\ \nonumber
U_{\alpha 3} & \gtrsim & \frac{(1 +\sqrt{1- B^{0}
(\nu _\alpha)})}{2}
:\mbox{close to one}.
\end{eqnarray}

Therefore, there are four possibilities.
\begin{enumerate}
\item
Region I. ~~~~Small~~ $ {\mid  U_{e3}\mid}^2$ 
~ and ~ ${\mid U_{\mu 3}\mid}^2$,
\item
RegionII.~~~~ Small~~ ${ \mid U_{e3}\mid}^2$
~ and ~ large 
${\mid U_{\mu 3}\mid}^2$,
\item
Region III.~~Large ~~$\mid U_{e3}\mid ^2$
~and~ small
$\mid U_{\mu 3}\mid$,
\item
Region IV. ~Large $\mid U_{e3}\mid ^2 $
~and~ $\mid U_{\mu 3}\mid ^2$,
\end{enumerate}
where $large$ means close to one but less than or
equal to one.
First, it is obvious that the region IV is not allowed simply 
because of the  unitarity condition.
The region III is also excluded because 
with  $\mid U_{e3}\mid ^2$ obtained from the 
oscillation data,
the solar neutrinos are depleted by only 
10 percents ,which is
in sharp contrast to what has been observed. Thus, we are left with
two regions only. To be more specific, in the model under
consideration, the survival probability of the solar
neutrinos ($\nu_{e}$) is given by \cite{giunti1}
\begin{equation}
P(\nu_{e} \rightarrow \nu_{e}) =
( 1- |U_{e3}|^2)^{2}
 P^{(1,2)} 
(\nu_{e} \rightarrow \nu_{e}) + |U_{e3}|^4,
\end{equation}
where 
$ P^{(1,2)} 
(\nu_{e} \rightarrow \nu_{e})$ is the survival
probability due to the mixing between the first and
the second generations. If the parameter $|U_{e3}|^2$
is large, we have,  for all values  of the neutrino
energy,
$ P 
(\nu_{e} \rightarrow \nu_{e}) \gtrsim 0.92 $,
which is not compatible with the results of the current 
solar neutrino experiments.  
It is worth commenting that the case I is consistent with
the standard mass hierarchy 
\begin{equation}
\langle m(\nu_e)\rangle  \ll \langle m(\nu_\mu)\rangle
\ll \langle m(\nu_\tau)\rangle,
\end{equation}
whereas the case II corresponds to an inverted hierarchy
\begin{equation}
\langle m(\nu_e) \rangle \ll \langle m(\nu_\tau)\rangle
\ll \langle m (\nu_\mu)\rangle.
\end{equation}
Although it is quite unnatural, this possibility is not
excluded in this analysis.
 
\subsection {Region I}
 The region I is consistent with the standard
 mass hierarchy of the $effective$ masses of the
 weak eigenstates.
We will consider the limits
 on the oscillation amplitude $A(\nu_\mu \rightarrow 
 \nu_e)$. If we simply take the negative results of BNL E776
 \cite{BNL}
 and KARMEN \cite{KARMEN} presented in the two generation
 form, the limits on the oscillation amplitude 
 $ A(\nu_{\mu} \rightarrow \nu_{e})$ 
 would be the same as those obtained in the
 two generation analysis.
 If the data are interpreted in this way, some part of
 the allowed region in the $\Delta m^2 - \mbox{sin}
 ^2(2\theta)$ plot obtained by
 the positive result of the LSND experiment
 is not ruled out by the previous negative
 results.
 However, as emphasized earlier, we can
 find additional constraints on 
 $ A(\nu_{\mu} \rightarrow \nu_{e})$ 
 since this oscillation channel is 
 related to other oscillations such as 
 $\nu_\mu \rightarrow
 \nu_\tau$ and $\nu_e  \rightarrow \nu_\tau$.
 For example, we have 
 \begin{eqnarray}
 A(\nu_e \rightarrow \nu_\tau) & = & 4 \left| U
  _{e3}\right |^2 \left |U_{\tau3}\right |^2
  \simeq 4 \left | U_{e3}\right |,\\ \nonumber
 A(\nu_\mu \rightarrow \nu_\tau) & = & 4 \left | U
  _{\mu3}\right |^2 \left |U_{\tau3}\right |^2
  \simeq 4 \left | U_{\mu3}\right | .
 \end{eqnarray}
 In addition, we have
 \begin{equation}
 A(\nu_{\mu} \rightarrow \nu_{e}) 
 = 4 \left |U
  _{e3}\right |^2 \left |U_{\mu3}\right |^2
  \simeq
 \frac{
 A(\nu_e \rightarrow \nu_\tau)
 A(\nu_\mu \rightarrow \nu_\tau)}{4}.  
\end{equation}
The above relations provide us with much more
stringent constraints on the 
 $A(\nu_{\mu} \rightarrow \nu_{e})$
 than the  ones from direct two generation
 analyses or disappearance data.
Using the negative results of the $\nu_{e} \rightarrow
\nu_{\tau}$ and $\nu_{\mu} \rightarrow \nu_{\tau}$
oscillation experiments, we can obtain the forbidden
region in the $\Delta m^2 - A(\nu_{\mu}
\rightarrow \nu_{e})$
plot which turns out to be much wider than the simple
two generation result and the allowed region by the 
LSND experiment happens to be well within this 
forbidden region. That is, the LSND result is not consistent
with the previous negative oscillation experiments
 if the mass hierarchy is the standard one. 


\subsection{Region II}
This region can be a solution because of the quadratic
nature of the relation(16) in $\left|U_{\alpha i}
\right|^2$. In this region, 
$\left|U_{e3}\right|^2$
 and $ \left|U_{\tau3}\right|^2$
 are small and
$\left|U_{\mu3}\right|^2$ is close to one 
 (but less than or equal to one). This implies that
 the standard mass hierarchy of effective weak eigenstate masses
 is inverted, i.e.,
\begin{equation}
\langle m(\nu_e)\rangle  \ll \langle m(\nu_\tau)\rangle
\ll \langle m (\nu_\mu)\rangle.
\end{equation}
First, as in the case of the Region I, we have direct limits on
 $A(\nu_{\mu} \rightarrow \nu_{e})$
 from the negative results of the
 $\nu_{\mu} \rightarrow \nu_{e}$ experiment of
 BNL E776 \cite{BNL}, which yields limits on
 $|U_{e3}|^2$ since in this case
 $|U_{\mu3}|^2$ is close to one.
On the other hand, the positive indication of
the LSND experiment sets the bounds
\begin{equation}
 A^{-}(\nu_{\mu} \rightarrow \nu_{e}) \le
 A(\nu_{\mu} \rightarrow \nu_{e}) \le
 A^{+}(\nu_{\mu} \rightarrow \nu_{e}),
 \end{equation}
 where 
$ A^{-}(\nu_{\mu} \rightarrow \nu_{e})$
and 
$ A^{+}(\nu_{\mu} \rightarrow \nu_{e})$ are, respectively,
the bounds for fixed values of
$\Delta m^2$.
The bounds on $|U_{e3}|^2$ can be found
from Eq.(26).
Also,
 in this case we can find the following relation
\begin{equation}
 A(\nu_{\tau} \rightarrow \nu_{e}) = 
 A(\nu_{\mu} \rightarrow \nu_{e}) 
 A(\nu_{\mu} \rightarrow \nu_{\tau})/4
 \end{equation}
 which shows that the oscillation
 $\nu_{e} \rightarrow \nu_{\tau}$
 is  significantly suppressed. 
 In contrast to the previous case of the standard mass
 hierarchy, the 
 $A(\nu_{\mu} \rightarrow \nu_{e})$ is not constrained 
 any further by 
 other oscillation experiments. Therefore, the LSND result
 is not in contradiction with the previous experiments.
 
 \subsection{Neutrinoless $\beta$ $\beta$ Decay}
 We have shown above that in the framework of the model
 with the mass hierarchy,  the data on the 
 terrestrial oscillation experiments and
 the solar neutrinos indicate that the mixing
 parameter 
 $|U_{e3}|^2$ 
 is small. If massive neutrinos are indeed Majorana
 particles, this result has important
 implications on
 $(\beta \beta)_{0\nu}$ decay
 experiments.
The Dirac and Majorana nature of neutrinos can best be tested or
 answered in the experiment of neutrinoless double beta decays.
Even after heroic efforts by countless experimentalists, 
no 
such decay has so far been observed. This process is possible
only if neutrinos are Majorana. The Majorana neutrinos inherently
violate the lepton number by two units which is a necessary
condition for neutrinoless double beta--decay process to occur.
Its rate is proportional to,  in the absence of the right--
handed coulping, the square of  an effective mass 
 \cite{kim} \cite{samoil}
 \cite{doi} \cite{ejiri}
 \begin{equation}
 \langle m_{\nu} \rangle =
  | \sum _{i}  U^2 _{ei} m_i \eta_{i}|
\end{equation}
where $\eta_{i} = \pm 1$ is the $CP$  phase of the Majorana
neutrino $\nu_{i}$. So far no 
 $(\beta \beta)_{0\nu}$ decay has been observed,  
  setting  limits on $\langle m_{\nu}\rangle 
\lesssim 1 \mbox{eV}$ \cite{doi} \cite{ejiri} \cite{moe}.

Because of the assumed mass hierarchy of the mass 
eigenstates, we have, from Eq.(28),
 \begin{equation}
 \langle m_{\nu} \rangle \simeq
 \left|U_{e3}\right|^2 
 m_{3}
 \simeq 
 \left|U_{e3}\right|^2 
 {\sqrt{\Delta m^2}}.
\end{equation}
In the Region I, both  
 $\left|U_{e3}\right|^2$ 
and
$\left|U_{\mu3}\right|^2$ 
are small and 
 $\left|U_{e3}\right|^2$
 is constrained simply by the
 Bugey \cite{bugey} experiment
alone for the interval of the experiment
$10 ^{-1}{\mbox{eV}^2} \lesssim  \Delta  m^2 \lesssim
10^3~~ \mbox{eV}^2$. For example, 
for $\Delta m^2 \le ~ 5~ \mbox{eV}^2$, we have
\begin{equation}
 \langle m_{\nu} \rangle 
\lesssim 10^{-1} \mbox{eV},
\end{equation}
which implies that we need 
the sensitivity of the
next-generation experiment on
$(\beta \beta)_{0\nu}$ decay. 

On the other hand, 
 $\left|U_{e3}\right|^2$
is severely constrained, in the region II, by
 BNL E776 and LSND
 experiments. For a wide range of $\Delta m^2$,
we have
\begin{equation}
\langle m_{\nu} \rangle
\lesssim 10^{-2} \mbox{eV},
\end{equation}
implying that if the Region II turns out to be
the right region, the observation of
$(\beta \beta)_{0\nu}$ decay 
becomes a formidable, if not impossible, task.
We emphasize that
in this case the LSND results are not completely ruled out
by the previous oscillation experiments. 

We can entertain some other possibilities in the mass 
hierarchy besides the standard mass hierarchy assumed 
here \cite{giunti2}. For example, we can assume
\begin{equation}
m_1 \ll m_2 \simeq m_3.
\end{equation}
This scheme was recently considered in \cite{moha}.
In favor of such a scheme, there are some cosmological
arguments and astrophysical arguments concerning
the $r$-process production of heavy elements in the
neutrino-heated ejecta of supernova. In this case, it can be
shown \cite{giunti2} that
\begin{equation}
\langle m \rangle \simeq m_3.
\end{equation}
In this case, neutrinoless double beta decay experiments
 can directly probe the mass of $\nu_3$.

\section{\bf{Solar and Atmospheric Neutrinos}}
Perhaps, the most intriguing indication, at present,  for
 the massive
neutrino comes from the solar neutrinos and to lesser
extent from the atmospheric neutrinos.
The all four solar neutrino experiments, Kamiokande \cite{Kamio}
, GALLEX \cite{GALLEX}, SAGE \cite{SAGE} and
Homestake \cite{home},  have seen the deficit of the
expected rates based on the so-called
standard solar model \cite{bahcall}.
 We list the latest experimental results
together with the predictions of the standard solar model of
Bahcall and Pinsonneault \cite{bahcall}. 
\begin{enumerate}
\item{\bf{Kamiokande}:}
\begin{eqnarray}
 \Phi_{expt} & = & \frac{(2.89 \pm 0.2  \pm 0.35)\times 10^{6}}
{{cm^{2}sec}},\\ \nonumber
 \Phi_{theor} & = & \frac{(6.62 \pm 0.7)\times 10^{6}}
{{cm^{2}sec}}.
\end{eqnarray}
\item{\bf{GALLEX,~~SAGE:}}
\begin{eqnarray}
\Sigma _{expt} & = & 74 \pm 8 ~~\mbox{SNU}, \\ \nonumber
\Sigma _{theor} & = & 132 \pm 7~~  \mbox{SNU}.
\end{eqnarray}
\item{\bf{Homestake:}}
\begin{eqnarray}
\Sigma _{expt} & = & 2.55 \pm 0.25~~ \mbox{SNU}, \\ \nonumber
\Sigma _{theor} & = & 8.1 \pm 1.0 ~~  \mbox{SNU}.
\end{eqnarray}
\end{enumerate}
 We note that the amount of the deficit is different for
different experiments. 
It has also been demonstrated that the deficit is most prominent
for  the medium energy neutrinos which come mainly from
the $^7\mbox{Be}$ and CNO reactions. This feature is
 almost impossible to explain away with the changes 
in relevant parameters for the Sun. Therefore,
in addition to somewhat different amounts of the deficit
in the four experiments, one must explain the disappearance of
the medium energy neutrinos from  the Sun. 
This can be nicely done with the MSW effects \cite
{MSW}. 

The MSW effects are the results of a peculiar resonance
behavior of neutrinos in matter.
When neutrinos pass through a medium, they see a potential
$V$ due to their coherent interactions with the particles
in matter. Since this potential is  linearly proportional to
$G_{F}$, the effects can be large if matter
density is sufficiently large. 
Neglecting the contribution from 
neutral current interactions, 
which does not play any role in the discussion of
the MSW effect, the effective potential
 $V$ induces an increase (in the case of $\nu_{e}$
in the ordinary matter) 
of $2 E_{\nu} V$ in the  mass squared
of $m_{\nu_{e}}$,  which is  $m_1^{2} {\cos}^2 {\theta} +
m_{2}^{2} {\sin}^2{\theta}$ in vacuum. 
This is due to the fact that weak eigenstate
neutrinos see the effective potential, 
not the mass eigenstates. As a consequence, the equation of 
motion of neutrinos in matter  becomes Eq.(37) with Eq.(38)
with the one-one element replaced by  
a new matter value
 which is  $m_1^{2} {\cos}^2 {\theta} +
m_{2}^{2} {\sin}^2{\theta} + 2E_{\nu}V $. 
Diagonalizing the matrix in Eq.(38) with the above
modification, one can find how neutrino massess and mixing
angles are effectively modified in matter. Changes in them
are such that when  certain resonance conditions 
among $\Delta m^2$ and the vacuum mixing angle  are met
and at the same time changes in matter density 
 are smooth, $\nu_{e}$ 
is very efficiently converted into, in the case of
two generations of neutrinos, $\nu_{\mu}$.
This is called the adiabatic process. If the change in
density is not smooth, the conversion is not very efficient
and called the non--adiabatic process.
Whether the process is  adiabatic or not depends very crucially
on the neutrino energy. This is where the energy dependence
of the $\nu_{e}$ conversion probability comes in. We note here that
the above four experiments have different energy thresholds
( GALLEX and SAGE have the same threshold). 
 If the solar
 neutrinos are depleted according to the way the MSW effects
are in operation, one can find two possible sets of solutions
for $\Delta m^2$ and $\sin^2(2\theta)$. One is the small
angle solution, the other being the large angle solution.
It is often said in the literature that the fit  with small
angle is better but this conclusion is based on the two generation
analysis and the large angle solution is as good as the small
angle solution. In any case, we shall discuss the small
angle solution only in this article.
The 90 percent  C.L. solution of the all four experiments is given
by 
\begin{eqnarray}
\Delta m^2  =  m_{2}^2 - m_{1}^2 & \simeq & 10^{-5} \\ \nonumber
{\sin}^2(2\theta) & \simeq & 5 \times 10^{-3}.
\end{eqnarray}
Assuming the standard mass hierarchy  of $ m_{2} \gg m_{1}$,
the above gives
\begin{equation}
 m_{2} \simeq    3 \times 10^{-3} {\mbox{eV}},
\end{equation}
Identifying this mass with $0.09 m_{c}^2 / M_{R}$ as given by
the quadratic seesaw mechanism (Eq.(27)) with SUSY SU(5), 
we find $M_{R} \simeq 10^{11}$ GeV. Given $M_{R}$, one finds
$m_{1} \simeq 10^{-8} $ eV and $m_{3} \simeq 10$ eV.

The vacuum oscillation explanation of the solar  neutrino
puzzle is also possible but it is marginally 
successful  at present.
In particular,  the absence of the medium energy 
solar neutrinos is non-trivial to explain with the vacuum
oscillation although it is not
impossible.
In general, the vacuum oscillation solution yields
smaller neutrino masses than those determined by
the MSW effect solution.
 For example, it gives
$m_{2} \simeq 10^{-5} \mbox{eV} $.
 Therefore, the seesaw 
mechanism predictions for $m_1$ and $m_2$ are accordingly
smaller.
 
 The atmospheric neutrino problem  has recently been getting
serious. The atmospheric neutrinos
used to be unwanted backgrounds for the
proton decay experiments but they
themselves have become the subject of
 important study.  With an exception of the Frejus experiment,
Kamiokande, IMB, Soudan II and MACRO all see the muon
deficit.
In order to reduce uncertainties coming from those
of calculations, it is customary to consider the ratios of
ratios, $R \equiv ({\frac{N_{\mu}}{N_{e}}})_{expt}/
({\frac{N_{\mu}}{N_{e}}})_{theor}$, where
$N_{\mu(e)}$ is the number of muon (electron) events induced
by $\nu_{\mu} (\nu_{e})$.
 The following is the summary of the recent
experimental results.
\begin{eqnarray}
R(\mbox{Kamioka}) & = & 0.60 \pm 0.06 \pm 0.05 \\ \nonumber
R(\mbox{IMB}) & = & 0.56  \pm 0.04 \pm 0.04 \\ \nonumber
R(\mbox{Soudan II}) & = & 0.75  \pm 0.16 \pm 0.14  \\ \nonumber
R(\mbox{MACRO}) & = & 0.73  \pm 0.06 \pm 0.12  \\ \nonumber
R(\mbox{Frejus}) & = & 0.99  \pm 0.13 \pm 0.08.
\end{eqnarray}
 Currently,  the most popular interpretation of the
anomaly is that
assuming the experiments (with an exception of Frejus)  are right,
 $\nu_{\mu}$'s 
in the atmospheric neutrinos with the expected excess of 
 a factor of two compared to 
$\nu_e$,  are somehow being
depleted on the way to the detectors. 
This depletion  can be attributed to the oscillation
 of $\nu_{\mu}$
into $\nu_{\tau}$. This conclusion is based on the observation 
that while $\nu{e}$'s  do not have
enough length (due to
small $\Delta m^2$) to travel to  oscillate into $ \nu_{\mu}$
or $\nu_{\tau}$,  the distances for  $\nu_{\mu}$
to travel  are just enough (due to
large $\Delta m^2$)  so that $\nu_{\mu}$ can oscillate
partially into $ \nu_{\tau}$. If this 
interpretation turns out to be correct,
the following set of the parameters can explain
the observed deficits
\begin{eqnarray}
\Delta m_{3,2}^2 \equiv m_{3}^2 - m_{2}^2 
& \simeq & 10^{-2} \mbox{eV}^2 \\ \nonumber
{\sin}^2(2\theta) \simeq 1.
\end{eqnarray}
Again in the spirit of the mass hierarchy, it means
 that  $m_{3} \simeq 10^{-1}$ eV. This is in contradiction with
the LSND result( because 
the LSND gives $\Delta m_{2,1}^{2} \simeq 1 \mbox {eV}$,
which implies $m_{2} \simeq $ eV 
) and it also means that neutrinos can not
play  a  significant role as hot dark matter.
One must obviously wait for better data for the atmospheric
neutrinos in order to  draw a firm  conclusion one way or the other.  
We conclude by saying that the LSND issue and the atmospheric
neutrino problem can conclusively be settled by planned
long--baseline experiments such as Fermilab to
Soudan with $ L= 730$ Km, CERN to Gran Sasso with
$L= 720$ Km and KEK to SuperKamiokande with $L= 250$ Km.
The region to be explored by these
long--baseline experiments in
the $\Delta m^{2} - {\sin}^{2}(2 \theta)$ plot is 
precisely the region relevant to the LSND and the atmospheric
neutrino issues. 
\acknowledgments
The author would like to thank the Organizing Committee
for the Korea-Japan Joint Winter School for Particle
and Nuclear Physics, in particular Professor H.S. Song
of the Center for Theoretical Physics, Seoul National
University for the hospitality during the school. 
He also like to thank
 Department of Physics, Korea Advanced Institute 
 of Science and Technology for the  
 hospitality extended to him
 while this manuscript was 
prepared.


\begin{references}

\bibitem{kim}
C.W. Kim and A. Pevsner, "Neutrinos in Physics and Astrophysics",
Contemporary Concepts in Physics, Vol.8 (Harwood Academic, Gordon
and Breach, Chur, Switzerland,1993).
\bibitem{bohm}
F. Bohm and P. Vogel, "Physics of Massive Neutronos" (Cambridge
University Press, Cambridge,1987). 
\bibitem{pal}
R.N. Mohapatra and P.B. Pal,"Massive Neutrinos in Physics and
Astrophysics" (World Scientific, Singapore,1991).
\bibitem{LSND}
C. Athanassopoulos et al., Phys. Rev. Lett. $\bf{ D 75}$,
2650 (1995).
\bibitem{samoil}
S. M. Bilenky and S. T. Petcov, Rev. of Mod. Phys. $\bf{59}$, 671
(1987).
\bibitem{pse}
L.~Wolfenstein, Nucl. Phys. $\bf{B 186}$, 147 (1981).
\bibitem{seesaw}
M. Gell-Mann, P. Ramond and R. Slansky,in
"Supergravity", ed F. van Nieuwenhuizen
and D. Freedman, (North Holland, Amsterdam,1979)
p. 315; T. Yanagida, Proc. of the Workshop on Unified Theory
and the Baryon Number of the Universe, KEK,Japan,
1979.
\bibitem{RGE}
S. Bludman, D. Kennedy and P. Langacker, Phys. Rev.$\bf{D 45}$,
1810 (1992).
\bibitem{maki}
B. Pontecorvo, JETP, $\bf{6}$, 429 (1958); Z. Maki, M.  Nakagawa
and S. Sakata, Prog. Theor. Phys. $\bf{28}$, 870 (1962).
\bibitem{wave}
C. Giunti, C. W. Kim and U. W. Lee, Phys. Rev. $\bf{D 44}$, 3635 (1991).
\bibitem{weak}
C. Giunti, C. W. Kim and U. W. Lee, Phys. Rev. $\bf{D 45}$, 1231 (1992).
\bibitem{carlo}
C. Giunti, C. W. Kim,
J.A. Lee  and U. W. Lee, Phys. Rev. $\bf{D 48}$, 4310  (1993).
\bibitem{giunti1}
"Short Base-Line Neutrino Oscillations and
Neutrinoless Double Beta Decay in the
Framework of Three Neutrino Mixing and a Mass Heirarchy",
S.M. Bilenky, A. Bottino, C. Giunti and C. W. Kim,
to be published.
\bibitem{giunti2}
"Short Base-Line Neutrino Oscilations and $(\beta \beta)_{0\nu}$
-decay in Schemes with an Inverted Mass Spectrum", S.M. Bilenky,
C. Giunti, C. W. Kim and S. T. Petcov, to be published.  
\bibitem{doi}
M. Doi, T. Kotani and E. Takasugi, Prog. Theor. Phys.
Suppl. $\bf{83}$, 1 (1985).
\bibitem{ejiri}
H. Ejiri, Physics and Astrophysics of Neutrinos,
ed. by M. Fukugita and A. Suzuki (Springer--
Verlag, Berlin, 1994 ). 
\bibitem{moe}
See, for example, M. K. Moe, Nucl. Phys. B
(Proc. Suppl.) {\bf 38}, 36 (1995).
\bibitem{bugey}
B. Achkar et al., Nucl. Phys. B {\bf4} 34, 503 (1995).
\bibitem{CDHS}
F. Dydak et al., Phys. Lett. B {\bf134}, 281 (1984).
\bibitem{CCFR}
I.E. Stockdale et al., Phys. Rev. Lett.{\bf 52}, 1384 (1984).
\bibitem{BNL}
L. Borodovsky et al., Phys. Rev. Lett. {\bf68}, 274 (1992).
\bibitem{KARMEN}
B. Armburster et al., Nucl. Phys. B (Proc. Suppl.).
\bibitem{moha}
D.O. Caldwell and R.N. Mohapatra, Phys.Lett. B{\bf 354}, 371 (1995).
\bibitem{Kamio}
K.S. Hirata et al., Phys. Rev. $\bf{D 65}$, 2241 (1991);
Y. Suzuki, Talk presented  at the $6^th$ International
Workshop on Neutrino Telescopes, Venezia, March 1994. 
\bibitem{GALLEX}
GALLEX Collaboration, Phys. Lett.$\bf{ B 314}$, 445 (1993);
Phys. Lett.$\bf{B 377}$ (1994).
\bibitem{SAGE}
SAGE Collaboration, Proc. of the $27^th$ Int. Conf. on
High Energy Physics, Glasgow, July 1994.
\bibitem{home}
R. Davis Jr., Talk presented at the $6th$ International
Workshop on Neutrino Telescopes, Venezia, March 1994.
\bibitem{bahcall}
J. N. Bahcall and M.H.  Pinsonneault, Rev. Mod. Phys.
$\bf{64}$, 885 (1992). See also hep-ph/9505425.
\bibitem{MSW}
S.P.~Mikheyev and A. Yu. Smirnov, Sov. J. Nucl. Phys.
$\bf{42}$,913 (1985); L. Wolfenstein, Phys. Rev. $\bf{65}$,
2369 (1978).
\end{references}
\end{document}